# Quantum jumps of light recording the birth and death of a photon in a cavity

Sébastien Gleyzes[1], Stefan Kuhr[*,1], Christine Guerlin[1], Julien Bernu[1], Samuel Deléglise[1], Ulrich Busk Hoff[1], Michel Brune[1], Jean-Michel Raimond[1] & Serge Haroche[1,2]

**A microscopic system under continuous observation exhibits at random times sudden jumps between its states. The detection of this essential quantum feature requires a quantum non-demolition (QND) measurement[1-3] repeated many times during the system evolution. Quantum jumps of trapped massive particles (electrons, ions or molecules[4-8]) have been observed, which is not the case of the jumps of light quanta. Usual photodetectors absorb light and are thus unable to detect the same photon twice. They must be replaced by a transparent counter 'seeing' photons without destroying them[3]. Moreover, the light has to be stored over a duration much longer than the QND detection time. We have fulfilled these challenging conditions and observed photon number quantum jumps. Microwave photons are stored in a superconducting cavity for times in the second range. They are repeatedly probed by a stream of non-absorbing atoms. An atom interferometer measures the atomic dipole phase shift induced by the non-resonant cavity field, so that the final atom state reveals directly the presence of a single photon in the cavity. Sequences of hundreds of atoms highly correlated in the same state are interrupted by sudden state-switchings. These telegraphic signals record, for the first time, the birth, life and death of individual photons. Applying a similar QND procedure to mesoscopic fields with tens of photons opens new perspectives for the exploration of the quantum to classical boundary[9,10].**

A QND detection[1-3] realizes an ideal projective measurement which leaves the system in an eigenstate of the measured observable. It can therefore be repeated many times, leading to the same result until the system jumps into another eigenstate under the effect of an external perturbation. For a single trapped ion, laser-induced fluorescence provides an efficient measurement of the ion's internal state[5-7]. The ion scatters many photons on a closed transition. This fluorescence stops and reappears abruptly when the ion jumps in and out of a metastable shelving state, decoupled from the illumination laser. Quantum jumps have also been observed between states of individual molecules[8] and between the cyclotron motional states of a single electron in a Penning trap[4]. As a common feature, all these experiments use fields to probe quantum jumps in matter. Our experiment realizes for the first time the opposite situation in which the jumps of a field oscillator are revealed via QND measurements performed with matter particles.

We exploit single-photon resolved light shifts experienced by an oscillating dipole in the field of a high-Q cavity. This resolution requires a huge dipole polarizability, which is achieved only with very special systems, such as circular Rydberg atoms[10] or superconducting qubits[11,12] coupled to microwave photons. In our experiment, the measurement of the light shift induced by the field on Rydberg atoms is repeated more than a hundred times within the average decay time of individual photons.

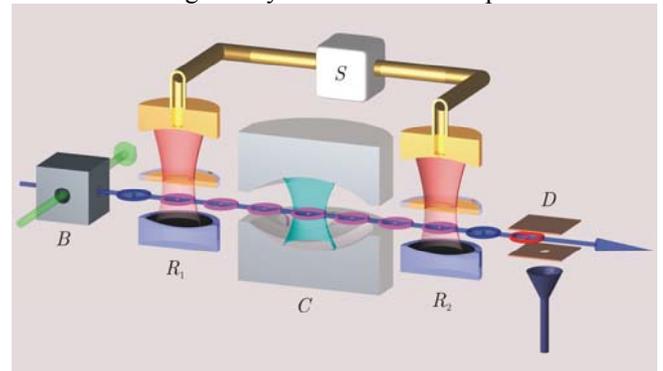

**Figure 1: Experimental set-up:** samples of circular Rydberg atoms are prepared in the circular state $g$ in box $B$, out of a thermal beam of Rubidium atoms, velocity selected by laser optical pumping. The atoms cross the cavity $C$ sandwiched between the Ramsey cavities $R_1$ and $R_2$ fed by the classical microwave source $S$, before being detected in the state selective field ionization detector $D$. The $R_1$-$C$-$R_2$ interferometric arrangement, represented here cut by a vertical plane containing the atomic beam, is enclosed in a box at 0.8 K (not shown) shielding it from thermal radiation and static magnetic fields.

The core of the experiment is a photon box (see Fig. 1), an open cavity $C$ made up of two superconducting niobium mirrors facing each other (Fabry Perot configuration)[13]. The cavity is resonant at 51.1 GHz and cooled to 0.8 K. Its damping time, as measured by the ring-down of a classical injected microwave field, is $T_c = 0.129 \pm 0.003$ s, corresponding to a 39,000 km light travel, folded in the 2.7 cm space between the mirrors. The QND probes are rubidium atoms, prepared in circular Rydberg states[10], travelling along the $Oz$ direction transverse to the cavity axis. They cross $C$ one at a time, at a rate of 900 s$^{-1}$ with a velocity $v$ = 250 m/s (see methods). The cavity $C$ is nearly resonant with the transition between the two circular states $e$ and $g$ (principal quantum numbers 51 and 50, respectively). The position-dependent atom-field coupling $\Omega(z)=\Omega_0 \exp(-z^2/w^2)$ follows the Gaussian profile of the cavity mode (waist $w$ = 6 mm). The maximum coupling, $\Omega_0/2\pi$ = 51 kHz, is the rate at which the field and the atom located at the cavity centre ($z$=0) exchange a quantum of energy, when the initially empty cavity is set at resonance with the $e$-$g$ transition[10].

If the atomic frequency is detuned from the cavity mode by $\delta/2\pi$ with $|\delta| \geq \Omega_0$, emission and absorption of photons by the probe atoms are suppressed due to the adiabatic variation of $\Omega(z)$ when the atom crosses the Gaussian cavity mode (see methods). The atom-field coupling results in shifts of the atomic and cavity frequencies[9]. The atomic shift depends on the field intensity and thus provides a QND information on the photon number $n$. Following a proposal made in Ref.14-15, our aim is to read this information by an interferometric method and to monitor the jumps of $n$ between 0 and 1 under the effect of thermal fluctuations and relaxation in the cavity,

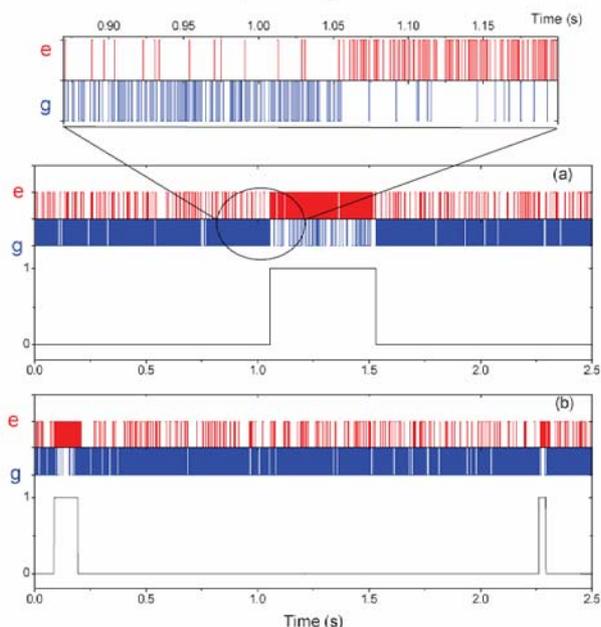

**Figure 2: Birth, life and death of a photon: a** QND Detection of a single photon. Red and blue bars show the raw signal, a sequence of atoms detected in $e$ or $g$ respectively (upper trace). The inset zooms into the region where the statistics of the detection events suddenly change, revealing the quantum jump from $|0\rangle$ to $|1\rangle$. The photon number inferred by a majority vote over 8 consecutive atoms is shown in the lower trace, revealing the birth, life and death of an exceptionally long lived photon. **b** Similar signals showing two successive single photons, separated by a long time interval with cavity in vacuum.

Before entering $C$, the atoms are prepared in a superposition of $e$ and $g$ by a classical resonant field in the auxiliary cavity $R_1$ (see Fig.1). During the atom-cavity interaction, this superposition accumulates a phase $\Phi(n,\delta)$. The atomic coherence at the exit of $C$ is probed by subjecting the atoms to a second classical resonant field in $R_2$, before detecting them in the state-selective counter $D$. The combination of $R_1$, $R_2$ and $D$ is a Ramsey interferometer. The probability for detecting the atom in $g$ is a sine function of the relative phase of the fields in $R_1$ and $R_2$. This phase is adjusted so that an atom is ideally found in $g$ if $C$ is empty ($n$ = 0). The detuning $\delta/2\pi$ is set at 67 kHz, corresponding to $\Phi(1,\delta) - \Phi(0,\delta) = \pi$. As a result, an atom is found in $e$ if $n$ =1. As long as the probability of finding more than one photon remains negligible, $e$ thus codes for the one-photon state, $|1\rangle$, and $g$ for the vacuum, $|0\rangle$. The probability for finding two photons in a thermal field at T = 0.8 K is only 0.3% and may be in first approximation neglected.

We first monitor the field fluctuations in $C$. Fig. 2a (upper part) shows a 2.5 s sequence of 2,241 detection events, recording the birth, life and death of a single photon. At first, atoms are predominantly detected in $g$, showing that $C$ is in $|0\rangle$. A sudden change from $g$ to $e$ in the detection sequence at $t$ =1.054 s reveals a jump of the field intensity, i.e. the creation of a thermal photon, which disappears at $t'$ = 1.530 s. This photon has survived 0.476 s (3.7 cavity lifetimes), corresponding to a propagation of about 143,000 km between the cavity mirrors.

The inset in Fig. 2a zooms into the detection sequence between times $t_1$ = 0.87 s and $t_2$ = 1.20 s, and displays more clearly the individual detection events. Imperfections reduce the Ramsey fringes contrast to 78%. There is a $p_{g|1}$ = 13% probability for detecting an atom in $g$ if $n$ = 1 and a $p_{e|0}$ = 9% probability for finding it in $e$ if $n$ = 0. Such misleading detection events, not correlated to real photon number jumps, are conspicuous in Fig. 2a and in its inset. To reduce their influence on the inferred $n$ value, we apply a simple error correction scheme. For each atom, $n$ is determined by a majority vote involving this atom and the last seven ones (see methods). The probabilities for erroneous $n$ = 0 ($n$ = 1) photon number assignments are reduced below $1.4\times10^{-3}$ ($2.5\times10^{-4}$) respectively per detected atom. The average duration of this measurement is $7.8\times10^{-3}$ s, i.e. $T_c/17$. The bottom part in Fig. 2a shows the evolution of the reconstructed photon number. Another field trajectory is presented in Fig. 2b. It displays two single photon events separated by a 2.069 s time interval during which $C$ remains in vacuum. By probing non-destructively the field in real time, we realize a kind of 'Maxwell demon' sorting out the time intervals during which the thermal fluctuations are vanishing.

Analyzing 560 trajectories, we find an average photon number $n_0 = 0.063 \pm 0.005$, slightly larger than $n_t = 0.049 \pm 0.004$, the thermodynamical value at the cavity mirrors temperature $0.80 \pm 0.02$ K. Attributing fully the excess photon noise to a residual heating of the field by the atomic beam yields an upper bound of the emission rate per atom of $10^{-4}$. This demonstrates the efficient suppression of atomic emission due to the adiabatic variation of the atom-field coupling. This suppression is a key feature making possible many repetitions of the QND measurement. Methods based on resonant phase shifts have much larger emission rates, in the $10^{-1}$ range per atom[3]. Non-resonant methods in which the detector is permanently coupled to the cavity[12] have error rates of the order of $\Omega_0^2/\delta^2$ and would require much larger $\delta/\Omega_0$ ratios to be compatible with the observation of field quantum jumps.

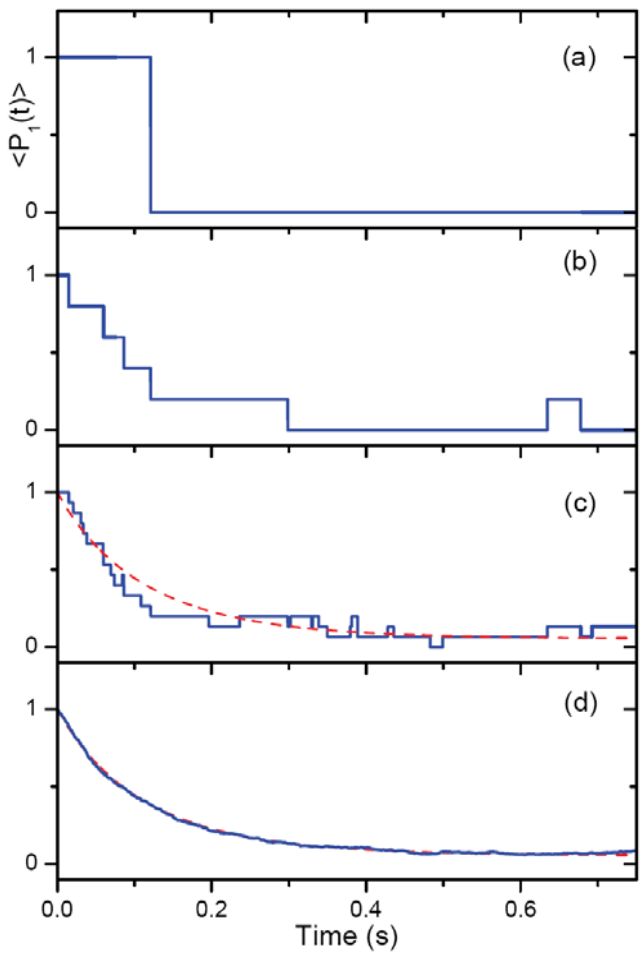

**Figure 3: Decay of the one-photon state: a** Measured value of $P_1=|1\rangle\langle 1|$ as a function of time, in a single experimental realization; **b-d** averages of 5, 15 and 904 similar quantum trajectories, showing the gradual transition from quantum randomness into a smooth ensemble average. Dotted red line in **c** and **d**: theoretical evolution of the probability of having one photon, $\langle P_1(t)\rangle$, obtained by solving the field master equation with the experimental values of $T_c$ and $n_0$.

In a second experiment, we monitor the decay of a single photon Fock state prepared at the beginning of each sequence. We initialize the field in $|0\rangle$ by first absorbing thermal photons with ~10 atoms prepared in $g$ and tuned to resonance with the cavity mode (residual photon number ~ 0.003±0.003). We then send a single atom in $e$, also resonant with $C$. Its interaction time is adjusted so that it undergoes half a Rabi oscillation, exits in $g$ and leaves $C$ in $|1\rangle$. The QND probe atoms are then sent across $C$. Fig. 3a shows a typical single photon trajectory (signal inferred by the majority vote) and Figs. 3b, c and d present the averages of 5, 15 and 904 such trajectories. The staircase-like feature of single events is progressively smoothed out into an exponential decay, typical of the evolution of a quantum average.

We have neglected so far the probability for finding 2 photons in $C$. This is justified, to a good approximation, by the low $n_0$ value. A precise statistical analysis reveals however the small probability of two photon events, which vanishes only at 0 K. When $C$ is in $|1\rangle$, it decays towards $|0\rangle$ with the rate $(1+n_0)/T_c$. This rate combines spontaneous ($1/T_c$) and thermally stimulated ($n_0/T_c$) photon annihilation. Thermal fluctuations can also drive $C$ into the two-photon state $|2\rangle$ at the rate $2n_0/T_c$ (the factor of 2 is the square of the photon creation operator matrix element between $|1\rangle$ and $|2\rangle$). The total escape rate from $|1\rangle$ is thus $(1+3n_0)/T_c$, a fraction $2n_0/(1+3n_0) \approx 0.10$ of the quantum jumps out of $|1\rangle$ being actually jumps towards $|2\rangle$.

In this experiment, the detection does not distinguish between $|2\rangle$ and $|0\rangle$. The incremental phase shift $\Phi(2,\delta) - \Phi(1,\delta)$ is $0.88\pi$ for $\delta/2\pi = 67$ kHz. The probability for detecting an atom in $g$ when $C$ is in $|2\rangle$ is ideally $[1-\cos(0.88\pi)]/2= 0.96$, indistinguishable from 1 within the experimental errors. Since the probability for $n > 2$ is completely negligible, the atoms precisely measure the projector $P_1=|1\rangle\langle 1|$, $e$ ($g$) coding for its eigenvalue 1 (0). Fig. 3d presents thus the decay of the ensemble average $\langle P_1(t)\rangle$, i.e. the probability for finding one photon in $C$. The theoretical expectation for $\langle P_1(t)\rangle$ (red dashed line in Fig. 3c and 3d) obtained by solving the field master equation[9,16] with the known values for $T_c$ and $n_0$ is nearly indistinguishable from the experimental data in Fig. 3d. Theory predicts - and experiment confirms - for $\langle P_1(t)\rangle$ a quasi-exponential decay with an initial slope corresponding to a time constant $T_c/(1+3n_0) = 0.109$ s, slightly shorter than $T_c =0.129$ s, the damping time of the average photon number.

Another analysis of the experimental data is provided by Fig. 4 which presents the histograms of the times $t$ of the first quantum jump after preparation of the field at $t = 0$ in $|1\rangle$ (circles) or $|0\rangle$ (squares). The histogram for $|1\rangle$ decays exponentially with the time constant $T_1 = 0.097 \pm 0.005$ s. The small difference with $T_c/(1+3n_0)$ is, within error bars, explained by wrong majority votes which can prematurely interrupt a one-photon detection sequence, with a negligible impact on $\langle P_1(t)\rangle$ (see methods),. The vacuum state is prepared by a first QND measurement of the thermal field in $C$ (first vote with majority in $g$). The detection ambiguity between $|0\rangle$ and $|2\rangle$ is then irrelevant. The histogram for $|0\rangle$ exhibits also an exponential decay, with $T_0 = 1.45 \pm 0.12$ s, whereas the expected lifetime is $T_c/n_0 = 2.05 \pm 0.20$ s. The difference is again mainly explained by the rate of false jumps, which affect most seriously the observed lifetime of long-lived states.

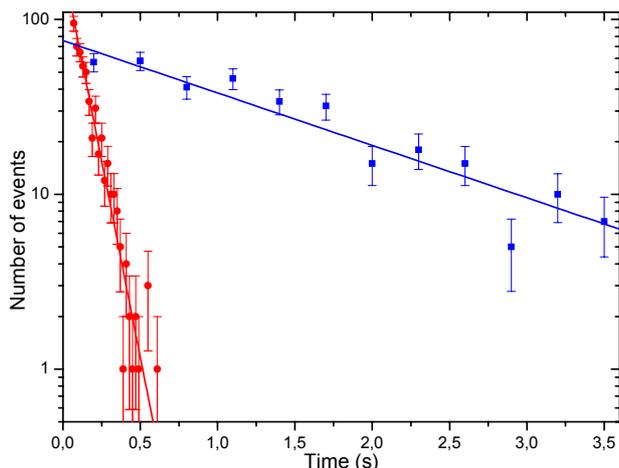

**Figure 4: Lifetimes of the one and zero photon states.** Histograms in log scale of the durations of the $|1\rangle$ (circles) and $|0\rangle$ (squares) states. The total number of events is 903 for $|1\rangle$ and 338 for $|0\rangle$. The error bars are statistical. The lifetimes, $T_1 = 0.097 \pm .005$ s and $T_0 = 1.45 \pm 0.12$ s are obtained from the linear fits (solid lines).

The atoms in this QND experiment are witnessing a quantum relaxation process whose dynamics is intrinsically not affected by the measurement. This is fundamentally different from experiments on micromaser bistability in which the dynamics of a coupled atom-field system exhibits jumps between two stable operating points[17]. Monitoring the photon number quantum jumps realizes an absolute radiation thermometer. The background photon number $n_0$, extracted from $\langle P_1 \rangle$ at thermal equilibrium, explains well the field states decay rates. Even though the two-photon states are not distinguished from the vacuum, their transient appearance with a small probability has an observable effect on the statistics. On single trajectories, however, the ambiguity between $|0\rangle$ and $|2\rangle$ can often be lifted by probabilistic arguments. In Fig. 2, for instance, the long time intervals in which $g$ is predominantly detected correspond certainly to vacuum, since their duration is much longer than the lifetime of $|2\rangle$, $T_c/(2+5n_0) = 0.057$ s.

This ambiguity is not a fundamental limitation of our QND scheme, which can be extended to monitor larger photon numbers[14,15,18]. By varying the settings of the Ramsey interferometer between probe atoms, we will be able to discriminate between different $n$ values. In the optimal setting[9,18], each detected atom extracts one bit of information about $n$. Ideally, this quantum analogue-digital converter pins down a state with a photon number $n$ between 0 and $N-1$ using only $\log_2(N)$ atoms. The first QND atom determines in this case the parity of $n$. Applied to a coherent state, this parity measurement projects the field into a Schrödinger cat state[15,19]. The photon number parity measurement will also allow us to reconstruct the Wigner function of the field in the cavity[20,21] and to follow its time evolution. The decoherence of Schrödinger cat states could be studied in this way[22], providing a direct observation of the evolution from quantum to classical behaviour in a mesoscopic system.

Finally, it is worth noting that, in this QND experiment, a single photon controls the state of a long sequence of atoms. The measurement amounts to a repetitive operation of hundreds of c-not gates[23] in which the same photon is the control bit (in its $|1\rangle$ or $|0\rangle$ state) and the successive atoms are the targets. This opens promising perspectives for multi-atom entanglement studies.

## METHODS

**Experimental set-up:** The principle of the circular Rydberg atom – microwave cavity set-up is presented in Refs. 9 and 10. A new superconducting mirror technology has been decisive in reaching very long photon storage times[13]. The mirrors are made of diamond-machined copper, coated with a 12 μm layer of niobium by cathode sputtering. The damping time $T_c$ is two orders of magnitude larger than that of our previous Fabry Perot cavities made up of massive niobium mirrors[10]. The cavity, whose mirrors have a toroidal surface, sustains two $TEM_{900}$ modes with orthogonal linear polarizations, separated in frequency by 1.2 MHz. The atomic transition is tuned close to resonance with the upper frequency mode by translating a mirror, using piezoelectric actuators. The atoms do not appreciably interact with the other mode. They enter and exit the cavity through large centimetre-sized ports, avoiding the stray electric fields in the vicinity of metallic surfaces. This ensures a good preservation of the atomic coherence. The Ramsey cavities must have a low $Q$ to minimize enhanced spontaneous emission of the atoms, and yet a well defined Gaussian mode geometry to preclude field leaking into $C$. To achieve these conflicting requirements, they are made of two parts coupled by a partly reflecting mirror (see Fig.1). The upper cavity, with $Q = 2 \times 10^3$ defines the mode geometry. It is weakly coupled to the lower one ($Q < 200$), crossed by the atoms. A QND detection sequence lasting 2.5 s consists of 35,700 atomic sample preparations, separated by 70 μs time intervals. The intensity of the lasers preparing the Rydberg states is kept low enough to limit the occurrence of two or more atoms per sample. This results in most samples being empty. On the average, we detect 0.063 single-atom events per sample. The average atomic detection rate is $r_a = 900$ s$^{-1}$. Each sample undergoes a classical $\pi/2$ pulse of 2 μs duration in $R_1$ and $R_2$. The first pulse prepares the atoms in $(|e\rangle + |g\rangle)/\sqrt{2}$. When $C$ contains $n$ photons, the uncoupled atom-cavity states $|e,n\rangle$ and $|g,n\rangle$ evolve into dressed states, shifted respectively, in angular frequency units, by $+([\delta^2 + (n+1)\Omega^2(z)]^{1/2} - \delta)/2$ and $-([\delta^2 + n\Omega^2(z)]^{1/2} - \delta)/2$. The difference between these frequencies, integrated over time ($t=z/v$) yields the phase shift $\Phi(n,\delta)$. Due to the smooth variation of $\Omega(z)$, the atom-cavity system follows adiabatically the dressed states. The final transition probability between $e$ and $g$ (obtained by numerical integration of the exact Schrödinger equation) is below $10^{-5}$ for $\delta/2\pi = 67$ kHz. Thus, $(|e\rangle + |g\rangle)/\sqrt{2}$ evolves at the exit of $C$ into $(|e\rangle + \exp[i\Phi(n,\delta)]|g\rangle)/\sqrt{2}$. When $\Phi(1,\delta) - \Phi(0,\delta) = \pi$, the Ramsey pulse in $R_2$ brings ideally the atom in g if $n=0$ and in e if $n=1$.

**Majority vote:** At each detection time, we determine the photon number by a majority vote, based on the outcomes of the last eight atomic measurements. In case of an equal 4/4 result, we retain the photon number from the preceding vote. This introduces a small hysteresis and reduces the rate of spurious jumps with respect to a simple majority vote with 7 or 9 atoms. The average duration of this measurement is $7.8 \times 10^{-3}$ s, resulting in a$\sim 3.9 \times 10^{-3}$ s delay between the occurrence of a quantum jump and its detection. We have determined by numerical simulations that a vote on 8 atoms is an optimal trade-off between errors and time resolution. The a priori probability of an error in a vote is given by the binomial law. With $p_{g|1}$=13%, we erroneously read 0 when there is 1 photon with a probability $\varepsilon_1 \sim (8!/3!5!)(0.13)^5(0.87)^3 = 1.4\times10^{-3}$. Similarly $p_{e|0}$=9% results in a false 1 reading with a probability $\varepsilon_0 \sim 2.5\times10^{-4}$. These errors are usually corrected after a time of the order of $7.8\times10^{-3}$ s, having thus a negligible impact on ensemble averages such as $\langle P_1(t)\rangle$, which evolve over a much longer time scale. They contribute however to an apparent increase of the $|1\rangle$ and $|0\rangle$ states decay rates. Computing the conditional probability for a vote to be erroneous while all the preceding ones are correct, we find additional decay rates of 0.61 s$^{-1}$ for $|1\rangle$ and 0.12 s$^{-1}$ for $|0\rangle$. Adding these figures to the theoretical decay rates of $|1\rangle$ and $|0\rangle$, we expect to get $T_1$ = 0.102 ±0.004 s and $T_0$ = 1.64±0.17 s.

**Acknowledgements.** We acknowledge funding by ANR, by the Japan Science and Technology Agency (JST), by the EU under the IP projects "QGATES" and "SCALA", and by a Marie-Curie fellowship of the European Community (S.K.). We thank P. Bosland, E. Jacques and B. Visentin for the niobium sputtering of the mirrors and Thomas Keating Ltd for providing Radar Absorbing material.

**Competing interests statement.** The authors declared no competing interests.

**Author Information.** S. Gleyzes and S. Kuhr contributed equally to this work. Correspondence and requests for materials should be addressed to M. Brune (brune@lkb.ens.fr).